\documentclass[prl,twocolumn,showpacs,preprintnumbers,amsmath,amssymb,citeautoscript]{revtex4-1}

\usepackage[dvipdfmx]{graphicx}
\usepackage{epstopdf}
\usepackage{amssymb, amsfonts, amsmath}
\usepackage{dcolumn}
\usepackage{bm}

\emergencystretch=\maxdimen
\usepackage{color}

\begin{document}

\title{Strongly correlated superconductivity in a copper-based metal-organic framework with a perfect kagome lattice}


\author{T.~Takenaka$^{1,{\ast}}$}
\author{K.~Ishihara$^{1,{\ast}}$}
\author{M.~Roppongi$^1$}
\author{Y.~Miao$^1$}
\author{Y.~Mizukami$^1$}
\author{T.~Makita$^1$}
\author{J.~Tsurumi$^1$}
\author{S.~Watanabe$^1$}
\author{J.~Takeya$^1$}
\author{M.~Yamashita$^2$}
\author{K.~Torizuka$^{2,3}$}
\author{Y.~Uwatoko$^2$}
\author{T.~Sasaki$^4$}
\author{X.~Huang$^5$}
\author{W.~Xu$^5$}
\author{D.~Zhu$^5$}
\author{N.~Su$^6$}
\author{J.-G.~Cheng$^6$}
\author{T.~Shibauchi$^{1,{\dag}}$}
\author{K.~Hashimoto$^{1,{\dag}}$}
\affiliation{
$^1$Department of Advanced Materials Science, University of Tokyo, Kashiwa, Chiba 277-8561, Japan\\
$^2$Institute for Solid State Physics, University of Tokyo, Kashiwa, Chiba 277-8581, Japan\\
$^3$Department of Physics, Nippon Institute of Technology, Miyashiro, Saitama 345-8501, Japan\\
$^4$Institute for Materials Research, Tohoku University, Aoba-ku, Sendai 980-8577, Japan\\
$^5$Beijing National Laboratory for Molecular Sciences, Key Laboratory of Organic Solids, Institute of Chemistry, Chinese Academy of Sciences, Beijing 100190, China\\
$^6$Beijing National Laboratory for Condensed Matter Physics and Institute of Physics, Chinese Academy of Sciences, Beijing 100190, China\\
$^{\ast}$These authors contributed equally to this work.\\
$^{\dag}$\rm{To whom correspondence should be addressed.\\
E-mail: shibauchi@k.u-tokyo.ac.jp (T.Sh.); k.hashimoto@edu.k.u-tokyo.ac.jp (K.H.)
}
}
\maketitle

{\bf
Metal-organic frameworks (MOFs), which are self-assemblies of metal ions and organic ligands, provide a tunable platform to search a new state of matter. A two-dimensional (2D) perfect kagome lattice, whose geometrical frustration is a key to realizing quantum spin liquids, has been formed in the $\bm{\pi}$-$\bm{d}$ conjugated 2D MOF [Cu$_{\bm{3}}$(C$_{\bm{6}}$S$_{\bm{6}}$)]$_{\bm{n}}$ (Cu-BHT). The recent discovery of its superconductivity with a critical temperature $\bm{T_{\rm c}}$ of 0.25\,kelvin raises fundamental questions about the nature of electron pairing. Here, we show that Cu-BHT is a strongly correlated unconventional superconductor with extremely low superfluid density. A nonexponential temperature dependence of superfluid density is observed, indicating the possible presence of superconducting gap nodes. The magnitude of superfluid density is much smaller than those in conventional superconductors, and follows the Uemura's relation of strongly correlated superconductors. These results imply that the unconventional superconductivity in Cu-BHT originates from electron correlations related to spin fluctuations of kagome lattice. 
}

\vspace{20pt}
\noindent
{\bf INTRODUCTION}
Metal-organic frameworks (MOFs), a subclass of coordination polymers with a nanoporous structure consisting of metal ions bridged by organic ligands\cite{Yaghi03}, are usually insulating because of the low degree of covalency of the metal-ligand bond. Recent substantial progress in the development of electrically conductive MOFs\cite{Talin14,Sun16,Ko18,Skorupskii20} may open up possibilities of applications in energy storage and chemical sensing, such as batteries\cite{Wada18}, thermoelectric devices\cite{Erickson15}, and chemiresistive sensors\cite{Campbell15}. However, most conductive MOFs are semiconducting, and thus the realization of delocalization of electrons, that is, band transport in MOFs is highly challenging.

A promising route toward metallic MOFs is the through-bond approach, which can be achieved by improving the covalency of the metal-ligand bond\cite{Sun16}. Under this strategy, highly conductive MOFs have been synthesized in two-dimensional (2D) layered frameworks\cite{Kambe13,Kambe14,Huang15,Dou17} composed of transition metal ions, such as Ni$^{2+}$ and Cu$^{2+}$, and multidentate organic ligands, such as benzenehexathiol (BHT), where the strong overlap of the $d$ orbitals of the metal ions and the $p$ orbitals of the organic ligand leads to delocalization of electrons with significantly higher conductivity, compared to conventional MOFs. However, in most cases, charge transport properties of such conductive 2D MOFs are dominated by thermally-activated conduction, indicative of incomplete bulk metallicity due to impurities or randomness coming from grain boundaries\cite{Kambe13,Kambe14,Huang15,Dou17}.

Among them, copper-BHT complex with a formula of [Cu$_3$(C$_6$S$_6$)]$_n$ (Cu-BHT) exhibits extremely high conductivity at room temperature ($\sim$2,500\,S$\cdot$cm$^{-1}$) (Ref.\,\onlinecite{Huang18}). This material consists of stacked $\pi$-$d$ conjugated 2D nanosheets with Cu$^{2+}$ ions and BHT ligands, forming a perfect kagome lattice of Cu$^{2+}$ with $S=1/2$ spins (see Fig.\,1, A to C). In Cu-BHT, in contrast to conventional MOFs with relatively large nanopores, the Cu ions and the BHT ligands are connected in an extremely dense manner (Fig.\,1A). This leads to a large hybridization of the $d$ orbitals of the Cu ions and the $\pi$ orbitals of the BHT ligands, resulting in an electronic structure having bands crossing the Fermi level, as confirmed by band structure calculations based on the lattice structure derived from x-ray diffractions\cite{Huang18}. This unique structure of Cu-BHT and recent substantial improvement of sample quality have enabled to realize metallization in Cu-BHT and, unexpectedly, even superconductivity at around 0.25\,K has been reported\cite{Huang18}. Note that to the best of our knowledge, this is the first report of superconductivity in coordination polymers.

\begin{figure}[h]
\begin{center}
\includegraphics[width=1\linewidth]{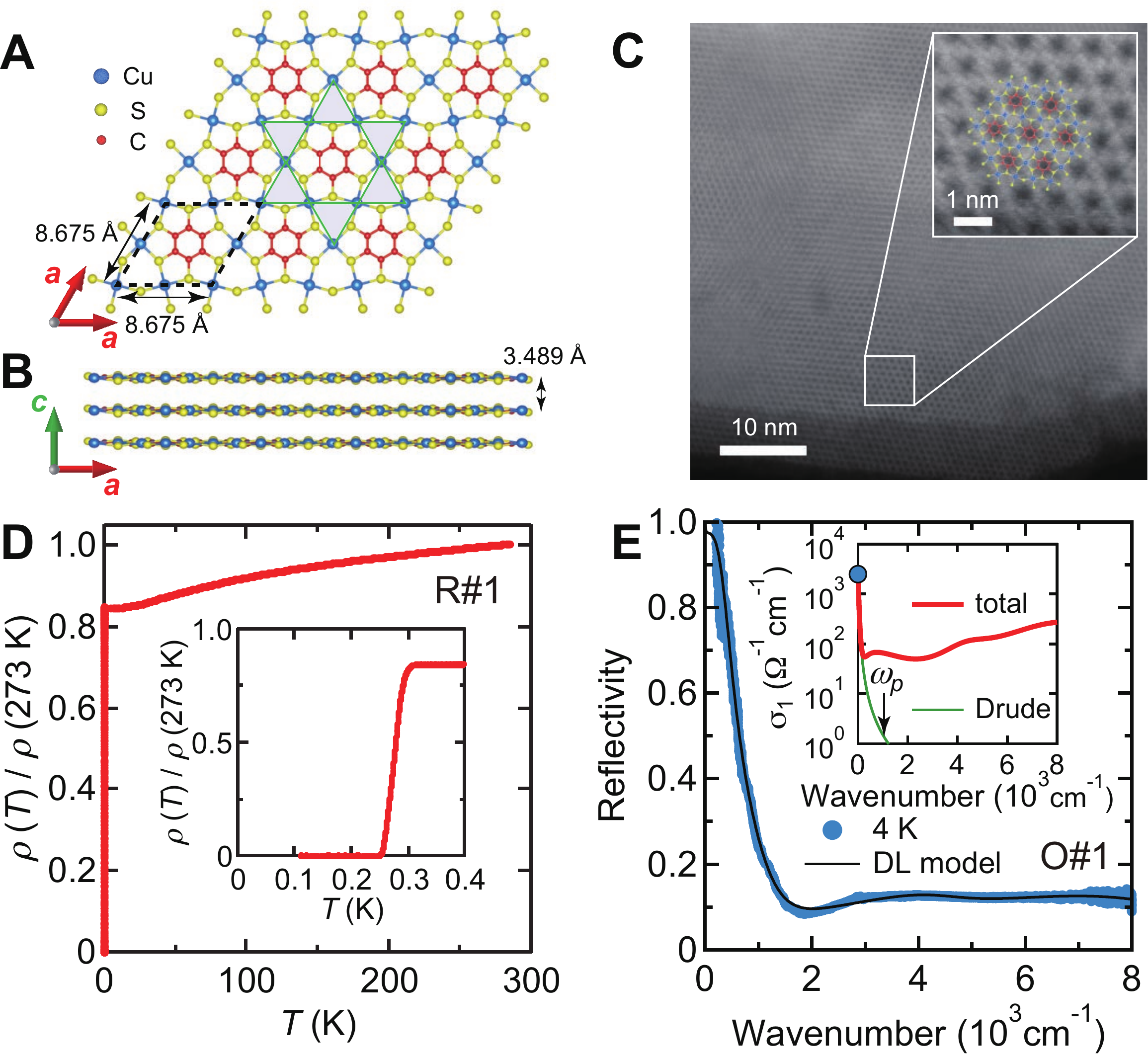}
\end{center}
\noindent
\caption{
{\bf Crystal structure of Cu-BHT and its transport and optical properties.}
({\bf A} and {\bf B}) Crystal structure of Cu-BHT viewed along the $c$ direction (A) and viewed from the side (B). The lattice parameters, $a=8.675$\,$\rm{\AA}$ and $c=3.489$\,$\rm{\AA}$, were determined by the powder x-ray diffraction pattern\cite{Huang18}. The black dotted parallelogram in (A) indicates the in-plane unit cell. The green shaded area indicates the perfect kagome lattice of Cu$^{2+}$ ions.
({\bf C}) High-resolution transmission electron microscopy image of Cu-BHT. The top right inset displays the amplified image with higher magnification focused on the square area in the main panel.
({\bf D}) Temperature dependence of the in-plane resistivity in Cu-BHT. The inset shows the low-temperature data below 0.4\,K. A sharp superconducting transition is observed at around 0.25 to 0.3\,K. 
({\bf E}) Optical reflectivity spectra (blue circles) in Cu-BHT measured at 4\,K. The black solid line is the fit to the Drude-Lorentz (DL) model. The inset shows the optical conductivity spectra (red line) obtained from the optical reflectivity (see the Supplementary Materials). The green line represents the Drude term. The blue circle in the inset shows the value of the dc conductivity obtained from the standard four-probe method at 4\,K ($\sigma_1(0)=1/\rho=2,500$\,ohm$^{-1}$cm$^{-1}$), which is consistent with $\sigma_1(0)$ obtained from the optical measurements. The plasma frequency $\omega_{\rm p}$ was estimated to be $\sim$880\,cm$^{-1}$, at which $\sigma_1(\omega)$ of the Drude term becomes almost zero.}
\end{figure}

The superconducting transition temperature $T_{\rm c}$ of Cu-BHT is quite low, and its pairing mechanism has been considered in the framework of the conventional weak-coupling Bardeen-Cooper-Schrieffer (BCS) theory based on electron-phonon coupling from first-principles calculations\cite{Zhang17}. However, its 2D perfect kagome lattice of $S=1/2$ spins is a longstanding desire in material chemistry and physics to realize quantum spin liquids, and its relation with high-$T_{\rm c}$ superconductivity has been widely discussed\cite{Anderson73,Anderson87,Balents10,Savary16}. Recent heat capacity and magnetic susceptibility measurements in Cu-BHT (Ref.\,\onlinecite{Huang18}) have revealed the absence of any long-range magnetic order down to 50\,mK and the presence of strong spin fluctuations related to the kagome lattice. Strongly enhanced spin fluctuations can promote unconventional (non-BCS) superconducting pairing, which has been observed in strongly correlated electron systems, such as high-$T_{\rm c}$ cuprates, iron-pnictides, organics, and heavy-fermions\cite{Keimer15,Si16,Powell11,Gegenwart08}. It is therefore quite important to experimentally determine whether the superconductivity in Cu-BHT has conventional or unconventional nature. Here, we report on a comprehensive study on the superconducting properties of Cu-BHT via transport, optical, and magnetic penetration depth measurements, which reveals that contrary to the theoretical proposal\cite{Zhang17}, Cu-BHT is a new member of strongly correlated unconventional superconductors, possibly originating from electron correlations enhanced by geometrical frustration of the kagome lattice.

\vspace{20pt}
\noindent
{\bf RESULTS}\\
\noindent
{\bf Transport and optical properties of Cu-BHT}

Highly crystalline samples of Cu-BHT with a thin-film structure studied here were prepared via the liquid-liquid interface reaction (see the Supplementary Materials)\cite{Huang15,Huang18}. Figure\,1D shows the temperature dependence of normalized in-plane dc resistivity $\rho$ in Cu-BHT, demonstrating a metallic behavior ($d\rho/dT>0$) down to low temperatures, accompanied by a clear superconducting transition at around 0.25\,K (see the inset of Fig.\,1D). Note that  we have measured multiple samples of Cu-BHT, all of which show a clear superconducting transition, but $T_{\rm c}$ is somewhat different among samples (see fig.\,S1). This could be related to the unconventional superconductivity of Cu-BHT, as discussed later.

The metallic behavior in Cu-BHT has been also confirmed in optical spectroscopy measurements. Figure\,1E shows the optical reflectivity spectra $R(\omega)$ of Cu-BHT. In general, in metals, the so-called Drude response can be observed in $R(\omega)$, in which two characteristic features are detected\cite{Dressel02}; one is a high reflectivity value close to unity in the low-energy region, and the other is the plasma edge given by $\omega_{\rm p}=\sqrt{e^2 n_{\rm{3D}}/(\epsilon_0 m^{\ast})}$ (where $n_{\rm{3D}}$ and $m^*$ are the 3D carrier density and effective mass of conducting electrons, respectively), above which $R(\omega)$ starts to saturate. Both features can be clearly seen in $R(\omega)$ of Cu-BHT; as wavenumber approaches zero, $R(\omega)$ goes to unity, whereas $R(\omega)$ shows a saturation at around 1000 to 2000 cm$^{-1}$ (see Fig.\,1E). The overall spectra of the reflectivity can be well fitted to the Drude-Lorentz model, from which we have extracted the corresponding optical conductivity spectra $\sigma_1(\omega)$ including a Drude term, as shown in the inset of Fig.\,1E (see the Supplementary Materials). These metallic features observed both in the dc limit and at finite frequencies provide strong evidence for the realization of band transport in Cu-BHT.

\vspace{20pt}
\noindent
{\bf Upper critical fields of Cu-BHT}

Having established that band transport is realized in Cu-BHT, we next discuss the superconducting nature of Cu-BHT. To this end, first we measured the upper critical field $ H_{\rm c2}$. Figure\,2 (A and B) shows the temperature dependence of in-plane resistivity in magnetic fields perpendicular ($H_{\perp}$) and parallel ($H_{\parallel}$) to the in-plane direction, respectively. The superconducting state is maintained up to higher magnetic fields for the $H_{\parallel}$ configuration than for the $H_{\perp}$ configuration. Figure\,2C depicts the temperature dependence of the perpendicular and parallel upper critical fields, $H_{\rm{c2}\perp}(T)$ and $H_{\rm{c2}\parallel}(T)$. For both directions, $H_{\rm c2}$ is well described by the Werthamer-Helfand-Hohenberg (WHH) model\cite{Werthamer66}. The observed anisotropy parameter $\gamma\equiv H_{\rm{c2}\parallel}(0)/H_{\rm{c2}\perp}(0)\sim1.5$ is comparable to those of iron-based superconductors such as SrFe$_2$(As$_{1-x}$P$_x$)$_2$ (Ref.\,\onlinecite{Yeninas13}) and LiFeAs (Ref.\,\onlinecite{Khim11}) with cylindrical Fermi surfaces ($\gamma = 1.4$ and $1.3$, respectively), indicating the quasi-2D nature of superconductivity in Cu-BHT. Note that the out-of-plane coherence length of Cu-BHT is estimated to be $\xi_{\perp} = 27$ nm, which is much shorter than the thickness of the superconducting region of the Cu-BHT samples studied here ($\sim2.5\,\mu$m), meaning that the quasi-2D superconductivity of Cu-BHT originates from the electronic structure itself, not from the thin-film effect.

\begin{figure}[h]
\begin{center}
\includegraphics[width=1\linewidth]{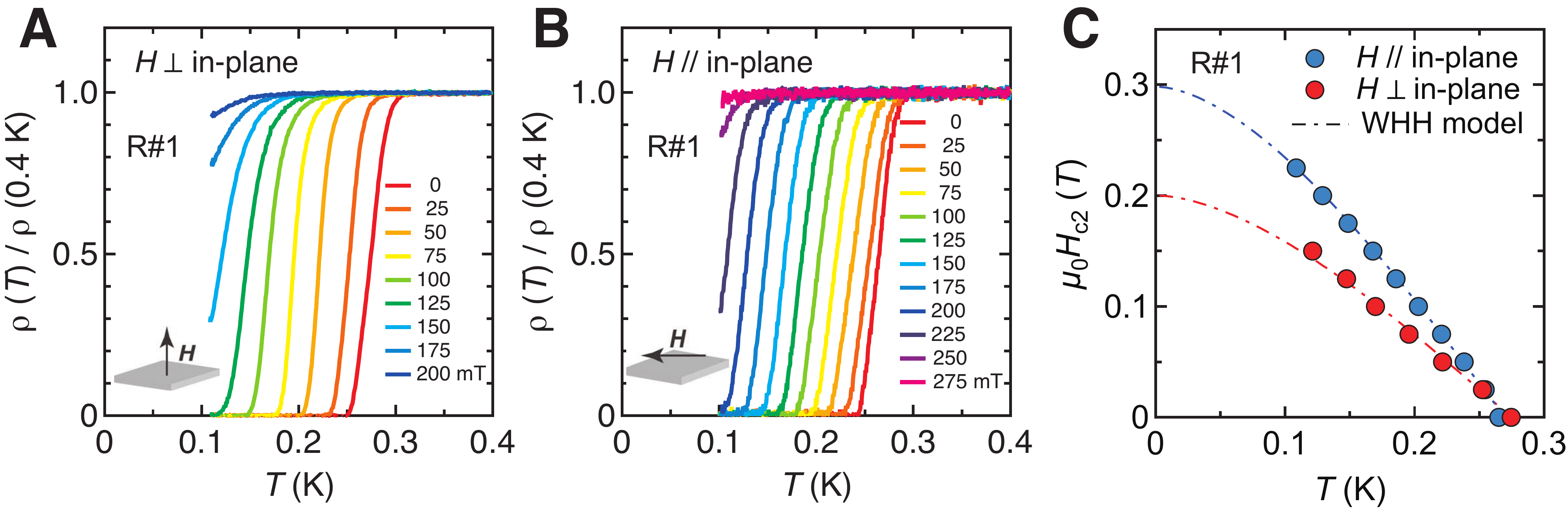}
\end{center}
\noindent
\caption{
{\bf Anisotropy of the upper critical fields of Cu-BHT.}
({\bf A} and {\bf B}) Temperature dependence of the in-plane resistivity in magnetic fields applied perpendicular (A) and parallel (B) to the in-plane direction.
({\bf C}) Temperature dependence of the upper critical fields for the $H_{\parallel}$ configuration (blue circles) and the $H_{\perp}$ configuration (red circles). Here, we defined $\mu_0 H_{\rm c2}$ as a field at which the resistivity becomes 50$\%$ of the normal-state value. The dashed lines represent the WHH model\cite{Werthamer66}. The anisotropy of $H_{\rm c2}$ is estimated to be 1.5, and the in-plane and out-of-plane coherence lengths were estimated to be $\xi_{\parallel} = 41$\,nm and $\xi_{\perp} = 27$\,nm, respectively, through the relations, $H_{{\rm c2}\perp}(0)=\Phi_0/(2\pi\xi_{\parallel}^2)$ and $H_{{\rm c2}\parallel}(0)=\Phi_0/(2\pi\xi_{\parallel}\xi_{\perp})$, where $\Phi_0=2.07\times10^{-15}$\,Wb is the flux quantum.}
\end{figure}

\vspace{20pt}
\noindent
{\bf Superconducting gap structure of Cu-BHT}

An important question concerning the newly found quasi-2D superconductivity in Cu-BHT is what is the interaction that glues the electrons into Cooper pairs. Conventional phonon-mediated pairing leads to a superconducting gap opening all over the Fermi surfaces, while unconventional pairing mechanisms, such as spin fluctuations, can lead to an anisotropic gap with nodes where the superconducting gap becomes zero\cite{Sigrist 91}. Such a nodal structure has been observed in $d$-wave superconductors such as cuprates, heavy fermions, and organic conductors, where the low-energy superconducting quasiparticle excitations remain finite even at low temperatures.

There are several physical quantities sensitive to the superconducting quasiparticle excitations. Among them, magnetic penetration depth $\lambda$, which is one of the most fundamental properties of superconductors\cite{Prozorov06}, is a sensitive probe of the low-energy quasiparticle excitations, and it directly relates with the 3D carrier density of superconducting electrons $n_{\rm{3D}}$ through the relation $\lambda^{2}(0)=m^{\ast}/(\mu_0 e^2 n_{\rm{3D}})$, where $\lambda(0)$ is the magnetic penetration depth at absolute zero, and $m^{\ast}$ is the effective mass of the superconducting carriers. In this study, we measured the magnetic penetration depth of Cu-BHT down to 40\,mK by using a tunnel diode oscillator (TDO) in a dilution refrigerator (see the Supplementary Materials). Figure\,3A shows the total frequency shift accompanied by the superconducting transition, which is directly related to the magnetic susceptibility $\chi$ and magnetic penetration depth $\lambda$. A clear drop in the frequency shift has been observed at 0.2 to 0.25\,K, which confirms the superconducting Meissner state, evidencing bulk superconductivity in Cu-BHT. A relatively broad superconducting transition may be attributed to the thin-film structure of Cu-BHT; the magnetic penetration depth near $T_{\rm c}$ becomes longer than the thickness of the superconducting region of the sample, which prevents a full superconducting shielding, resulting in the broad superconducting transition{\cite{Kim18}. The importance of phase fluctuations in superconductors with small superfluid density has been discussed in quasi-2D superconductors\cite{Emery95}, which may also broaden the superconducting transition. Since Cu-BHT has a 2D structure with extremely low superfluid density discussed later, the effect of phase fluctuations near $T_{\rm c}$ may affect the superconducting transition width.

Next we focus on the low-temperature part of the magnetic penetration depth measurements. The absolute value of $\lambda(0)=\sqrt{m^{\ast}/(\mu_0 e^2 n_{\rm{3D}})}$ is estimated to be $\sim1.8\,\mu$m from the plasma frequency $\omega_{\rm p}=\sqrt{e^2 n_{\rm{3D}}/(\epsilon_0 m^{\ast})}\approx 880$ cm$^{-1}$ obtained in the optical measurements at 4\,K through the relation $\lambda(0)=c/\omega_{\rm{p}}$ (here, $c=1/\sqrt{\epsilon_0\mu_0}$ is the velocity of light), which is comparable to the typical thickness of the superconducting region of the present Cu-BHT samples ($\sim2.5\,\mu$m). In general, when the thickness $t$ of a superconductor is comparable to the magnetic penetration depth $\lambda$, we need to consider the so-called Pearl length $\Lambda_{\rm eff} = \lambda^2 /t_{\rm eff}$ [where $t_{\rm eff} = 2\lambda \tanh (t/2\lambda)$] as the 2D screening length\cite{Pearl64,Chen14}.  By considering such a thin-film effect, we obtained the magnetic penetration depth from the measured frequency shift (for details, see the Supplementary Materials). Figure\,3B depicts $\delta\lambda(T)/\lambda(0)$ as a function of $T/T_{\rm c}$, where $\delta\lambda(T)\equiv\lambda(T)-\lambda(0)$. In conventional BCS superconductors with an isotropic full-gap structure, the temperature variation of the magnetic penetration depth is of an activated type at low temperatures\cite{Prozorov06}, $\delta\lambda(T)/\lambda(0) \approx \sqrt{\pi\Delta_0/(2k_{\rm B} T)}\exp[-\Delta_0/(k_{\rm B} T)]$ with $\Delta_0=1.76 k_{\rm B}T_{\rm c}$. Thus, $\delta\lambda(T)/\lambda(0)$ becomes almost $T$ independent at low temperatures. In stark contrast, $\delta\lambda(T)/\lambda(0)$ of Cu-BHT shows a steeper slope at low temperatures (see Fig.\,3B).

\begin{figure}[h]
\begin{center}
\includegraphics[width=1\linewidth]{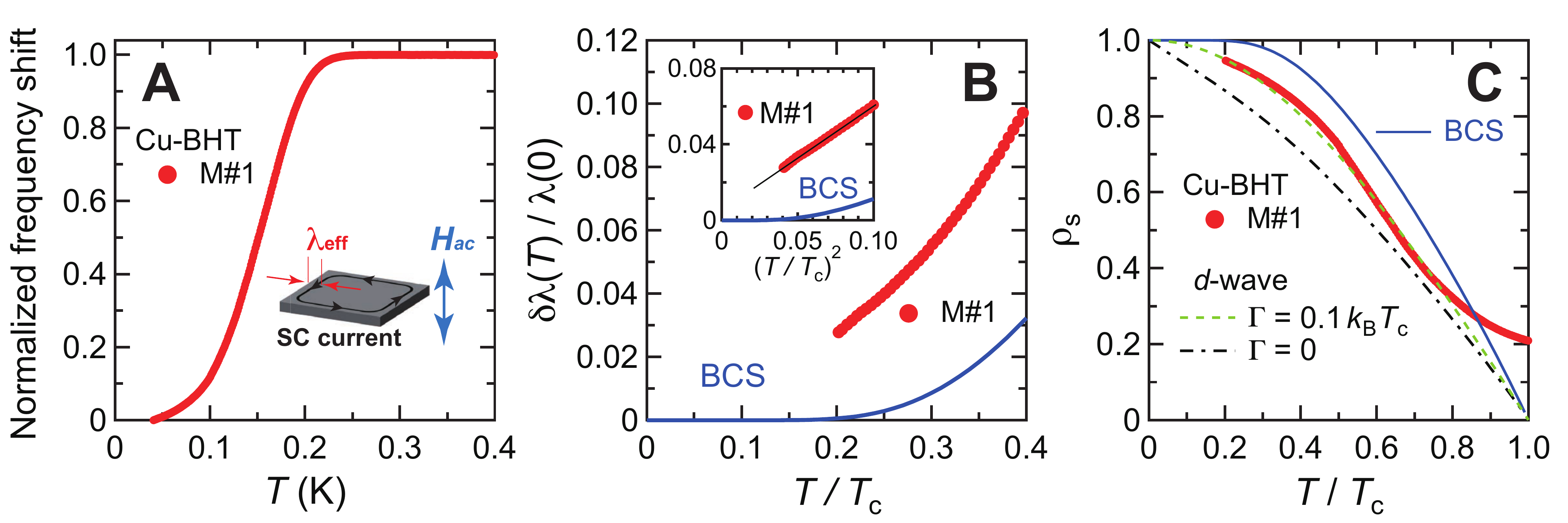}
\end{center}
\noindent
\caption{
{\bf Magnetic penetration depth and normalized superfluid density of Cu-BHT.}
({\bf A}) TDO frequency shift normalized by the total shift during the superconducting transition. The inset shows the schematic illustration of the superconducting screening current in an ac magnetic field.
({\bf B}) Change in the magnetic penetration depth normalized by $\lambda(0)$ (red circles) plotted against $T/T_{\rm c}$. The blue solid line represents the fully gapped behavior expected in the conventional $s$-wave (BCS) case ($\Delta_0=1.76k_{\rm B}T_{\rm c}$). Inset: The same data plotted against $(T/T_{\rm c})^2$.
({\bf C}) Normalized superfluid density plotted against $T/T_{\rm c}$. The blue solid line shows the behaviors expected in the $s$-wave (full gap with $\Delta_0=1.76k_{\rm B}T_{\rm c}$) case. The green (scattering parameter $\Gamma=0.1k_{\rm B}T_{\rm c}$) and black ($\Gamma=0$) lines represent the $d$-wave (line-node gap with $\Delta_0=2.14k_{\rm B}T_{\rm c}$) with and without impurities, respectively\cite{Sun95}. }
\end{figure}

In unconventional superconductors with line nodes in the gap such as $d$-wave cuprates, the low-temperature $\delta\lambda(T)$ is proportional to $T$ (Ref.\,\onlinecite{Hirschfeld93}). There are a few mechanisms that can affect the exponent $\alpha$ in the power-law temperature dependence of $\delta \lambda(T) \propto T^\alpha$. It has been discussed that the effects of non-magnetic impurity scattering\cite{Hirschfeld93} and quantum criticality\cite{Hashimoto13} tend to increase the exponent $\alpha$. In $d$-wave superconductors, the non-magnetic impurity effect\cite{Hirschfeld93} changes the low-temperature variation of $\lambda$ from $T$ to $T^2$, whereas in quantum critical superconductors with line nodes, a $T^{1.5}$ dependence is often found, which is related to the temperature-dependent mass renormalization\cite{Hashimoto13,Mizukami16,Nomoto13}. Thus, the low-temperature exponent $\alpha$ in the line-node case is expected as $1\le \alpha \le2$. As shown in the inset of Fig.\,3B, $\delta\lambda(T)/\lambda(0)$ follows a $T^2$ dependence. Thus, the obtained $T^2$ behavior is consistent with the presence of line nodes in the superconducting gap function with impurities. 

Next, we consider magnetic impurity effects that can change the temperature dependence of the magnetic penetration depth. If the normal state of a superconductor is paramagnetic, the measured magnetic penetration depth $\lambda_{\rm{m}}(T)$ is given by $\lambda_{\rm{m}}(T)=\lambda_{\rm{L}}(T)\sqrt{1+\chi_{\rm N}(T)}$ (Ref.\,\onlinecite{Cooper96}), where $\lambda_{\rm{L}}$ is the London penetration depth and $\chi_{\rm N}(T)=1+C/(T+\theta_{\rm{CW}})$ is the normal-state  susceptibility (here, $C$ is the Curie-Weiss constant and $\theta_{\rm{CW}}$ is the characteristic temperature for magnetic interaction). Since $\chi_{\rm N}(T)$ increases with decreasing temperature, $\lambda_{\rm{L}}(T)$ shifts downward from the value of $\lambda_{\rm{m}}(T)$ more largely at lower temperatures. Therefore, if the paramagnetic effect exists in Cu-BHT, $\lambda_{\rm{L}}(T)$ shows a steeper temperature dependence than $\lambda_{\rm{m}}(T)$ and the exponent of $\delta\lambda_{\rm{L}}(T)\propto T^{\alpha}$ becomes smaller than that of $\delta\lambda_{\rm{m}}(T)$. Such analysis results have been obtained in electron-doped cuprates\cite{Cooper96,Prozorov00-1} and iron-based superconductors\cite{Malone09}. Thus, even if we consider the paramagnetic effect for Cu-BHT, the exponent of $\delta\lambda_{\rm{L}}(T)\propto T^{\alpha}$ is expected to be smaller than 2, suggesting the presence of gap nodes in Cu-BHT.
We also note that in the presence of local magnetic impurities that act as free spins, conventional ($s$-wave) full-gap superconductivity can change to a gapless state, resulting in a $T^2$ dependence of $\delta\lambda(T)$ (Refs.\,\onlinecite{Prozorov06,Skalski64}). However, in our present experiment down to $\sim$40\,mK, no Curie term due to local magnetic impurities has been observed in the frequency change of the TDO.
The TDO measurements are very sensitive to the presence of local magnetic impurities, which can be detected as a low-temperature upturn in the frequency change described by the Curie law with $\theta_{\rm CW}\sim 0$. Such an upturn in the TDO technique has been observed in iron-based superconductors with a small amount of magnetic impurities on the order of 0.1\% in volume\cite{Hashimoto10}. Therefore, the possibility of gapless excitations due to magnetic impurities is unlikely in the present Cu-BHT case. 
 
One may also consider that strong quantum fluctuations as discussed below can break the Cooper pairs, leading to a nonexponential behavior of $\delta\lambda(T)$ even in a full-gap superconducting state. Since the magnitude of renormalization due to quantum fluctuations increases with decreasing temperature, the lower the temperature, the greater the effect of renormalization due to quantum fluctuations on the magnetic penetration depth\cite{Nomoto13}. As a result, in the case of line-node superconductors, the $T$-linear dependence of $\delta\lambda(T)/\lambda(0)$ changes to a $T^{1.5}$ dependence in a wide temperature range as mentioned above. Likewise, in the case of fully gapped superconductors, $\delta\lambda(T)/\lambda(0)$ is expected to show a flatter temperature dependence at low temperatures. This implies that in the full-gap case, power-law behavior due to strong quantum fluctuations cannot be expected.

The superconducting transition temperature of Cu-BHT is as low as 0.25 K, making it difficult to measure the magnetic penetration depth down to low enough temperatures to precisely determine the presence or absence of nodes in the gap. Therefore, for more detailed analysis of the superconducting gap structure,  we plotted the normalized superfluid density $\rho_{\rm s}(T)\equiv\lambda^2(0)/\lambda^2(T)$ as a function of $T/T_{\rm c}$ (see Fig.\,3C). The overall temperature dependence of $\rho_{\rm s}$ is consistent with that expected in $d$-wave superconductors with impurities\cite{Sun95} (green dashed line), evidencing the emergence of unconventional superconductivity with sign change of the superconducting order parameter in Cu-BHT. We note that a deviation from the theory near $T_{\rm c}$ may come from the thin-film effect and/or phase fluctuations near $T_{\rm c}$, discussed above. 

\vspace{20pt}
\noindent
{\bf DISCUSSION}

\begin{figure}[h]
\begin{center}
\includegraphics[width=0.8\linewidth]{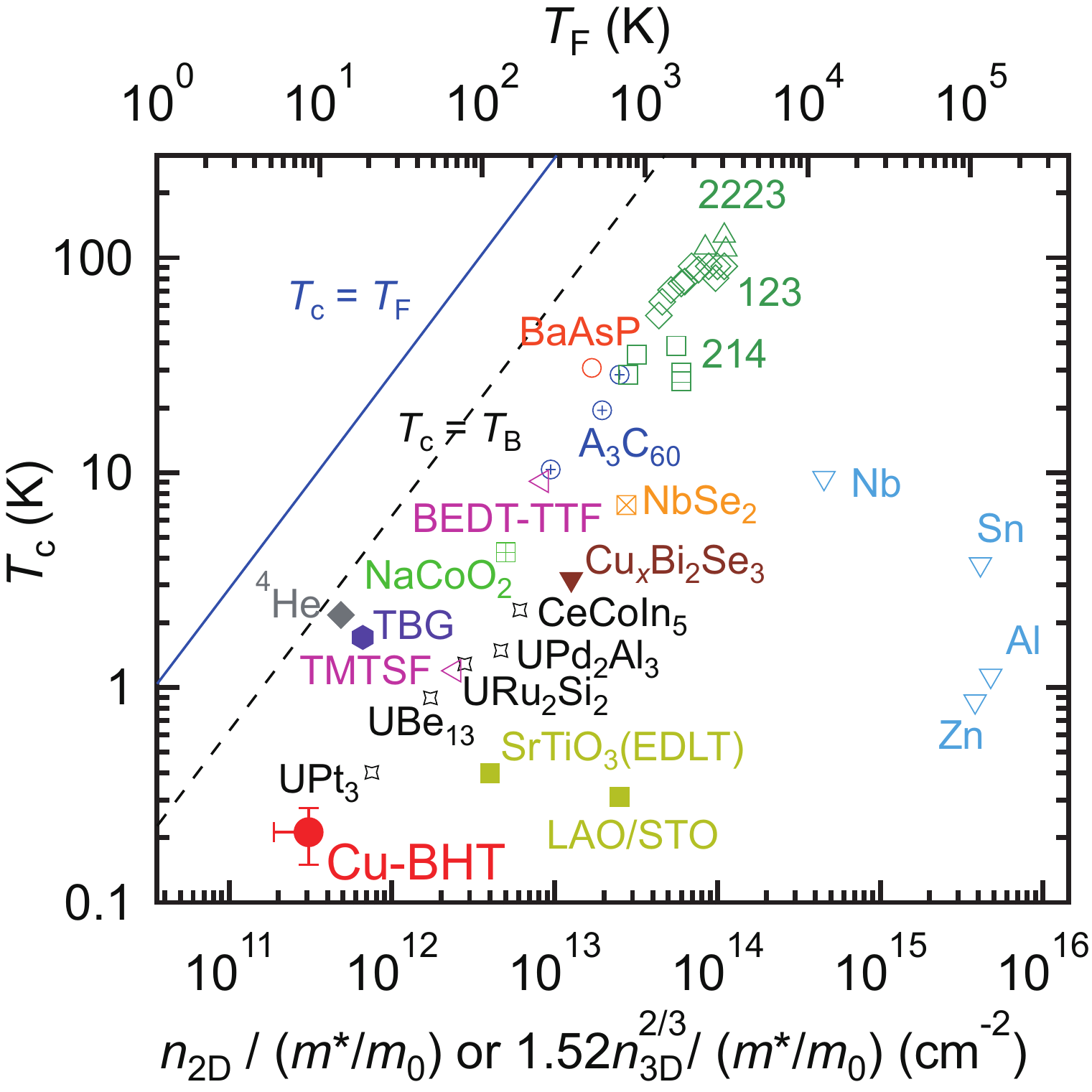}
\end{center}
\noindent
\caption{
{\bf Uemura plot.}
$T_{\rm c}$ is plotted against the effective superfluid density (bottom axis) given by $n_{\rm{2D}}/(m^{\ast}/m_0)$ for 2D systems and $1.52n_{\rm{3D}}^{2/3}/(m^{\ast}/m_0)$ for 3D systems, where $n_{\rm{2D}}$ is the carrier concentration within the superconducting planes for 2D systems, $n_{\rm{3D}}$ is the carrier concentration for 3D systems, and $m_0$ is the free electron mass. Here $n_{\rm{2D}}=n_{\rm{3D}}\times d$, where $d$ is the interlayer spacing of the superconducting planes in 2D systems. Note that $T_{\rm F}$ (top axis) is proportional to the effective carrier density $n_{\rm{2D}}$ through the relation $T_{\rm F}=\hbar^2\pi n_{\rm{2D}}/(k_{\rm B} m^{\ast})$. For Cu-BHT, the 2D formula was used (for details, see the Supplementary Materials). The error bar of $T_{\rm c}$ is determined by the sample dependence in the magnetic susceptibility measurements (see fig.\,S1B).
The error bar of $T_{\rm F}$$(=\frac{\hbar^2\pi d}{\mu_0 e^2 k_B}\lambda^{-2}(0)$) comes from the uncertainty of the absolute value of the zero-temperature magnetic penetration depth $\lambda(0)$. In this study, we evaluated $\lambda(0)$ from the plasma frequency $\omega_{\rm{p}}$ $=880\pm80\,\rm{cm}^{-1}$ measured at 4 K through the relation $\lambda(0)=c/\omega_{\rm{p}}$. Since the previous specific heat studies \cite{Huang18} point to an increase in the effective mass below 1\,K due to possible quantum fluctuations, the possibility that $T_{\rm F}$ decreases owing to the temperature variation of the effective mass below 1\,K should be taken into account, which is the main source of the error of $T_{\rm F}$. Therefore, we evaluated the error bar by considering that the electronic specific heat coefficient $\gamma_{\rm{el}}\propto m^{\ast}$ is enhanced below 1\,K from 15\,mJ$\cdot$mol$^{-1}$K$^{-2}$ above 1\,K to 40\,mJ$\cdot$mol$^{-1}$K$^{-2}$ at 0.2\,K.
The black dashed line is the Bose-Einstein condensation temperature for the ideal 3D boson gas. The blue solid line represents the line where $T_{\rm c}=T_{\rm F}$. Here, BEDTTTF and TMTSF stand for bis(ethylenedithio)tetrathiafulvalene and tetramethyltetraselenafulvalene, respectively, and LAO and STO represent LaAlO3 and SrTiO$_3$, respectively. TBG and EDLT stand for twisted bilayer graphene and electric double layer transistor, respectively.}
\end{figure}

To discuss the unconventional superconductivity of Cu-BHT in comparison with various superconductors, we constructed the so-called Uemura plot (an empirical relation between $T_{\rm c}$ and the effective Fermi temperature $T_{\rm F}$, see Fig.\,4), together with the results of various types of superconductors\cite{Uemura89,Uemura91,Uemura04,Hashimoto12,Cao18}. Here $T_{\rm F}$ is proportional to the effective 2D carrier density given by $n_{\rm{2D}}/(m^{\ast}/m_0)$ for 2D systems (see the Supplementary Materials). It is widely discussed that the ratio of $T_{\rm c}$ to $T_{\rm F}$ reflects the strength of the superconducting pairing interaction; in conventional weak-coupling BCS superconductors, such as Al and Sn, $T_{\rm c}/T_{\rm F}$ is quite low ($\sim10^{-5}$), whereas in strongly correlated superconductors, such as cuprates, iron-pnictides, organics, and heavy fermions, $T_{\rm c}/T_{\rm F}$ becomes high ($\sim10^{-2}$). As shown by the light blue triangles in the Uemura plot, in the conventional superconductors, only a tiny portion of electrons near the Fermi energy experience the superconducting gap, and $T_{\rm c}$ strongly varies in materials with similar superfluid densities depending on the superconducting strength. In sharp contrast, in the strongly correlated unconventional superconductors, $T_{\rm c}$ has a strong correlation with superfluid density, which is close to the linear relation expected for the Bose-Einstein condensation where the strong coupling between electrons leads to molecular-like bound pairs. What is notable here is that Cu-BHT is located on the trend line on which all the strongly correlated superconductors lie ($T_{\rm c}/T_{\rm F}=0.025$ in Cu-BHT). This result implies that the unconventional superconductivity in Cu-BHT originates from strong electron correlations like other strongly correlated electron systems. We also point out that the superfluid density in this system is extremely low comparable to heavy fermion systems, corresponding to the very low $T_{\rm c}\approx0.25$\,K, which places Cu-BHT on the bottom left corner in the Uemura plot.

A key question raised here is what is the mechanism of the strongly correlated unconventional superconductivity in Cu-BHT. As reported in the previous work\cite{Huang18}, the normal-state heat capacity $C$ and magnetic susceptibility show anomalous behaviors. The temperature dependence of $C/T$ below 1\,K follows $C/T \sim T^{-2/3}$, which is indicative of non-Fermi liquid behavior. This implies that the effective mass in the zero-temperature limit is significantly enhanced from the 4-K value, which pushes the superfluid density even smaller than that estimated from the optical spectroscopy (see the error bar in Fig.\,4). Such a temperature dependence of $C/T$ setting in at very low temperatures appears to be consistent with the extremely low effective Fermi temperature ($\sim8.5$\,K) in this system. 
The Curie-Weiss analysis of the magnetic susceptibility\cite{Huang18} also shows an effective moment of $\mu_{\mathrm{eff}}\sim 1.79 \mu_{\rm B}$ close to that expected for Cu$^{2+}$ with $S=1/2$ spins ($1.73\mu_{\rm B}$), with no sign of long-range magnetic ordering at least down to 2\,K despite a large magnitude of the Weiss temperature of $-1,400$\,K (Ref.\,\onlinecite{Huang18}).
These results suggest that the present system is close to a metallic quantum spin liquid state with strong quantum fluctuations arising from the geometrically frustrated kagome-lattice structure, as discussed in geometrically frustrated Kondo lattice systems\cite{Nakatsuji06,Tokiwa14}. Recent theoretical calculations\cite{Jiang17} have predicted the coexistence of frustrated local spins of Cu$^{2+}$ and itinerant electrons of $\pi$ orbitals in 2D MOF materials. Therefore, the unconventional pairing mechanism related to spin fluctuations in a metallic spin liquid state may be relevant in Cu-BHT. Our present findings may motivate experimental and theoretical studies on the relationship between unconventional superconductivity and quantum spin liquids. Considering the flexibility of designing crystal structures in MOFs, MOFs can provide a promising platform to study physical phenomena in condensed matter physics\cite{Yamada16,Dong16,Li17,Yamada17_1,Yamada17_2}.\\

\noindent
{\large \bf Materials and Methods}

\noindent
{\bf Sample preparation.} Highly crystalline samples of Cu-BHT were synthesized by the liquid-liquid interface reaction between BHT/chloroform and copper(II) nitrate/H$_2$O, as described in Ref.\,13. The typical lateral size of the samples is larger than $1\times1\,$mm$^2$, while the thickness is as small as a few micrometers. Cu-BHT films for the magnetic penetration depth measurements were cut into small pieces of samples with dimensions of about 350\,$\mu$m by 350\,$\mu$m.

\noindent
{\bf Electrical transport measurements.} The in-plane electrical resistivity was measured by the standard four-probe method in a dilution refrigerator down to 100\,mK. The electrical contacts were made on the surface using carbon paste. The applied current was reduced to less than 500\,nA to avoid Joule heating.

\noindent
{\bf Optical reflectivity measurements.} The optical reflectivity measurements were carried out with a Fourier transform microscope spectrometer in the range of 200 to 8,000\,cm$^{-1}$. In the far-infrared region (200 to 600\,cm$^{-1}$), a synchrotron radiation light source at BL43IR in SPring-8 was used. The optical conductivity was obtained by fitting the optical reflectivity to the Drude-Lorentz model [56]. The absolute value of the reflectivity was determined by comparison with a gold thin film evaporated on a glass plate, which was attached on the sample holder where the samples were fixed.

\noindent
{\bf Magnetic penetration depth measurements.} The temperature variation of the in-plane magnetic penetration depth $\delta\lambda(T)$ was measured by using the TDO technique operating at a resonant frequency of $\sim $14\,MHz in a dilution refrigerator down to $\sim40$\,mK (Ref.\,46). The sample was mounted on a sapphire rod with Apiezon N grease and inserted into a copper coil that is a part of the LC circuit. The shift in the resonant frequency $\delta f$ directly reflects the change in the magnetic penetration depth $\delta\lambda$. The samples were cooled slowly (with a rate less than 1.0\,K/min) to avoid introducing cracks into samples.\\

\noindent
{\bf SUPPLEMENTARY MATERIALS}\\
\noindent
section\,S1. Sample dependence of the superconducting transition temperature\\
\noindent
section\,S2. Estimation of $\lambda(0)$ from optical reflectivity measurements\\
\noindent
section\,S3. Thickness of superconducting region\\
\noindent
section\,S4. Derivation of magnetic penetration depth from frequency shift in thin films\\
\noindent
section\,S5. Uemura plot\\
\noindent
fig.\,S1. Sample dependence of the superconducting transition temperature\\
\noindent
fig.\,S2. Upper critical fields\\
\noindent
fig.\,S3. Thin film structure of Cu-BHT\\
\noindent
fig.\,S4. Frequency shift of the TDO in Cu-BHT\\
\noindent
table\,S1. Optical parameters of the DL model

\vspace{20pt}
\noindent
{\large \bf References and Notes}

\noindent{\bf Acknowledgements:} We thank N.\,Hosono, T.\,Kitao, K.\,Kanoda, and E.\,-G.\,Moon for discussions, and Y.\,Ikemoto and T.\,Moriwaki for technical assistance. Far-infrared reflectivity measurements using a synchrotron radiation light source were performed at SPring-8 with the approval of the Japan Synchrotron Radiation Research Institute (2018B0073). A part of the work was carried out under the Visiting Researcher's Program of the Institute for Solid State Physics, the University of Tokyo. {\bf Funding:} This work was supported by Grants-in-Aid for Scientific Research (KAKENHI) (Nos.\,JP20H02600, JP20K21139, JP19H00649, JP19K22123, JP19H01848, JP19K21842, JP19H00648, JP18H01853, JP18KK0375, JP18J11307), Grant-in-Aid for Scientific Research on Innovative Areas ``Quantum Liquid Crystals'' (No.\,JP19H05824), Grant-in-Aid for Scientific Research for Transformative Research Areas (A) ``Condensed Conjugation'' (No.\,JP20H05869) from Japan Society for the Promotion of Science (JSPS), CREST (No.\,JPMJCR19T5) from Japan Science and Technology (JST), National Key R\&D Program of China (Grant Nos.\,2017YFA0204701, 2018YFA0305700), the National Science Foundation of China (22071256, 21790051, 12025408, 11921004, 11834016, 11874400), the Beijing Natural Science Foundation (Z190008), and the Strategic Priority Research Program and Key Research Program of Frontier Sciences of the Chinese Academy of Sciences (XDB250000000, XDB33000000, QYZDB-SSWSLH013) as well as the CAS Interdisciplinary Innovation Team (JCTD-2019-01). {\bf Author contributions:} T.Sh. and K.H. conceived the project. T.T., K.I., M.R., Y.Mi., Y.Mi., T.Sh. and K.H. performed the magnetic penetration depth measurements. K.I., T.M., J.Ts., S.W., and J.Ta. carried out the SEM characterizations. T.T. K.I., M.Y., K.T., Y.U., and K.H. performed the electrical transport measurements. T.T., T.Sa. and K.H. performed the optical reflectivity measurements. X.H., W.X., D.Z., N.S., and J.G.C. carried out sample preparation. K.H. prepared the manuscript with inputs from T.T., K.I., and T.Sh. All authors discussed the experimental results. {\bf Competing interests:} The authors declare that they have no competing interests. {\bf Data and materials availability:} All data needed to evaluate the conclusions in the paper are present in the paper and/or the Supplementary Materials. Additional data related to this paper may be requested from the authors.\\

\end{document}